\documentstyle[11pt]{article}
\begin{document} 

\renewcommand{\thefootnote}{\alph{footnote}}
\begin{center}
{\Large\bf The Structure of Flux-tubes in SU(2)\\[1ex]}

A.M.~Green\footnotemark[1]$^,$\footnotemark[3],
C.~Michael\footnotemark[2]$^,$\footnotemark[4],
P.S.~Spencer\footnotemark[1]$^,$\footnotemark[5]\\

$\,^a$Research Inst. for Theoretical Physics, University
of Helsinki, Finland\\
$\,^b$Theoretical Physics Division, Dept. of Math. Sciences, 
University of Liverpool, Liverpool, UK.
\end{center}
\setcounter{footnote}{3}
\footnotetext{email: {\tt green@phcu.helsinki.fi}}
\setcounter{footnote}{4}
\footnotetext{email: {\tt cmi@liv.ac.uk}} 
\setcounter{footnote}{5}
\footnotetext{email: {\tt paddy.spencer@parallax.co.uk}; current address: Parallax Solutions Ltd., Stonecourt, Siskin Drive, Coventry CV3 4FJ, UK.}
\setcounter{footnote}{0}
\renewcommand{\thefootnote}{\arabic{footnote}}
 
\begin{abstract}  The spatial distribution of the  action and energy in
the colour fields of  flux-tubes is studied in lattice $SU(2)$ field
theory for static quarks at separations up to 1 fm. Special attention is
paid to the structure of the colour fields associated with an excited
flux-tube with E$_u$ symmetry. We compare our results with hadronic
string and  flux-tube models.  Sum rules are used to extract  generalised
$\beta$-functions, to describe the expected colour field  behaviour and
to cross check the methods used.
                  
\medskip

\end{abstract}

\section{Introduction}

Confinement in QCD is a non-perturbative phenomenon and lattice gauge
theory  techniques are well suited to its study. One of the simplest
manifestations  of confinement in pure glue QCD is the potential energy
$V(R)$ between static  quarks at separation $R$ which increases with
$R$.  To investigate this  in detail, it is possible to probe the spatial
distribution of the colour  fields around such static quarks. This
exploration of the nature of the  flux between static quarks has a long
history. Typical points of interest are the transverse extent of the 
flux-tube and  the nature of the colour fields (i.e. electric  or
magnetic). A hadronic string approach gives a reasonable  description of
the interquark potential at large separation. Here we  explore this
further by investigating the distribution of colour fields  for gluonic
excitations of this potential and comparing our results with models  for
the flux-tube in this case also.

On a lattice the technique to explore the colour field distributions is 
to measure the correlation of a plaquette $\Box$ [here defined as 
$\Box={1 \over 2} {\rm Tr} (1-U_{\Box})$] with the generalised Wilson
loop $W(R,T)$  that creates the static quark-antiquark at separation
$R$. See Fig.~1 for an illustration.  Locating the plaquette at $t=T/2$,
in the limit $T \to \infty$  the following expression isolates the
contribution from the colour field at position ${\bf r}$  from the
plaquette  oriented in the  $\mu,\ \nu$ plane. 
\begin{equation}  \label{fmnT} 
 f_R^{\mu \nu}({\bf r})=\left[{\langle
W(R,T)
  \Box^{\mu \nu}_{\bf r}\rangle} 
-\langle W(R,T)\rangle \langle \Box^{\mu \nu}\rangle 
\over {\langle W(R,T)\rangle} \right]
 \end{equation} 

These contributions are related in the naive continuum limit to the mean
squared fluctuation of the  Minkowski colour fields by 
 \begin{equation}
\label{fmn} 
 f_R^{ij}({\bf r})\rightarrow  \frac{a^4}{\beta}B^2_k({\bf r}) \quad
{\rm with} \ i,\ j,\ k\   {\rm cyclic \ \ \ and} \quad
f_R^{i4}({\bf r})\rightarrow -\frac{a^4}{\beta}E^2_i({\bf r}). 
\end{equation}
 \noindent We shall wish to distinguish electric and magnetic colour
fields with different orientations with respect to the interquark
separation axis (longitudinal -- taken as 1-axis) and adopt  the notation
that  
 \begin{equation} \label{ae3}  
 {\cal E}_L =  f^{41},\; {\cal E}_T =  f^{42,43},\; {\cal B}_T = 
f^{12,13},\; {\cal B}_L =   f^{23,32}. 
 \end{equation}  
 \noindent It will be convenient to discuss combinations corresponding
naively to  the action ($S$) and energy ($E$) densities of the gluon
field. These  are given by 
 \begin{equation} \label{ae}
S({\bf r})=-({\cal E}_L + 2  {\cal E}_T  + 2  {\cal B}_T + {\cal B}_L )
 \end{equation} 
 \noindent and 
\begin{equation}
 \label{ae2} 
 E({\bf r})=E_L({\bf r})+2E_T({\bf r})=
-({\cal E}_L - {\cal B}_L )- 2 ( {\cal E}_T  -  {\cal B}_T ).
 \end{equation}  
 \noindent Note that since the colour field contributions come from  a
cancellation between the value in a static potential and that in the
vacuum, either sign  is possible.

 The colour field distributions are quite difficult to measure on  a
lattice because the correlation of Eq.~(\ref{fmnT}) involves delicate
cancellations. Recent  results~\cite{bali} and~\cite{hay} have
concentrated on the simpler  case of SU(2) colour in order to achieve
sufficient statistical accuracy. It is expected  that the salient
features of confinement in SU(2) are very similar  to those in the
realistic case of SU(3). Here we focus our attention on the colour field
distributions  at $\beta=2.4$ for which the lattice spacing $a \approx
0.1$ fm  should be sufficiently small for our purposes. The  action and
energy densities are measured on a lattice by using  plaquettes of area 
$a^2$ and hence do not have a continuum limit as $a \to 0$. To  achieve
a continuum limit would require to measure energy and action  by using
loops of fixed physical size rather than plaquettes. In the following, 
we quote our results in lattice units with the normalisation as given
above  in Eq.~(\ref{fmnT}).

 Different strategies have been used to improve the signal to noise in
the  lattice determination of the colour fields.  Here we focus on one
method, as described in Sect.  3,   which is to create operators which
produce ground state potentials with  very little contamination of
excited states. As described in the next section, we are able to achieve
this using fuzzed links and a variational approach.  Because we wish to
study sum rules which involve sums over all spatial positions, we do 
not employ multi-hit techniques such as those used in Ref.~\cite{bali}.
We  do not use, either, the proposal of Ref.~\cite{hay} to 
treat $\langle \Box^{\mu \nu}\rangle $ in Eq.~(\ref{fmnT}) as a quasi
local average of the plaquette in the correlation since, although it 
reduces the noise, it also modifies the signal. 

 Lattice sum rules have been derived~\cite{cmsr,rothe,cm} which relate
the  sum over all space of the plaquette correlation to expressions
involving  the potential $V(R)$ and generalised $\beta$-functions. These
sum rules  have been used in an attempt to extract such
$\beta$-functions~\cite{gluelump,wup}. Here we consider a larger set  of
sum rules and   we make a careful evaluation of   these sum rules to
extract the generalised $\beta$-functions and to validate  our methods.
This is described in Sect.  4. A preliminary account  has been given 
in Ref.~\cite{letter}. 

 We present our results for the spatial distributions of the colour
fields  in Sect.  5. One feature that goes beyond previous work, is
that we  also measure the colour field distribution in the lowest lying
gluonic excitation of the static quark potential. This lowest gluonic
excitation  has non-zero angular momentum about the interquark axis and
plays a role in hybrid meson spectroscopy.  Hadronic string models
predict such excited modes and agreement with  the energy levels
determined  on a lattice has been found~\cite{phm,pm}. Here we are able
to explore the spatial  distribution of this excited gluonic mode to make
more detailed comparisons with models. We have also explored a higher gluonic 
excited state: the first excited state with the symmetry of the ground state. 
In this case we have been unable to extract unambiguously the colour 
fields associated with this state so we do not report on it further here.

\section{Operators for static potentials}

To explore the colour field distributions around static quarks at 
separation $R$, we need to find efficient lattice operators to  create
such states. Efficient here means having a small contamination  of
excited states while still allowing a good signal to noise for
correlations of interest. We build on the experience of studying the
corresponding potentials~\cite{phm} and use an iterative fuzzing
prescription to create the  spatial paths $P_i$ of length $R$, where 
different levels of iteration of fuzzing are represented by $i$. In
detail, we use a number of recursive iterations  of replacing each
spatial link by the projected sum of (\mbox{4 $\times$ straight link $+$
4 spatial U-bends}) and then  multiply these fuzzed links to make the
spatial path needed.   The correlation of spatial path $P_i$ at time $t$
and $P_j$ at time  $t+T$ then gives the generalised Wilson loop
$W_{ij}(T)$ - see Fig.~1. We use a $16^3 \times 32$ lattice at
$\beta=2.4$ for this  study with the $SU(2)$ gauge group. For many
purposes $SU(2)$ colour  provides an excellent test bed for QCD studies.
In this work the scale  as set by the string tension corresponds to a
lattice spacing $a \approx 0.6$ GeV$^{-1} \approx 0.12$ fm.  The data
sample used comes from a comprehensive study of  80 blocks (60 blocks
for $R\le 3$) of 125 measurements separated by 4 update sweeps (3 
over-relaxation plus  one heatbath). Error estimates use a full
bootstrap analysis of these blocks.

The fuzzed link operator $P_i$ to create a static quark and antiquark at
separation $R$ with colour field in a state of a given lattice symmetry
will have an expansion in terms of the  eigenstates of the transfer
matrix
 \begin{equation}
|R\rangle=c_0|V_0\rangle+c_1|V_1\rangle+\dots
 \end{equation}
with the measured correlation of a generalised Wilson loop given by
 \begin{equation}
W(R,T)=\langle R_0|R_T\rangle = 
c_0^2 e^{-V_0 T} \left(1 + h^2 + \dots \right)
 \end{equation}
where 
 \begin{equation}
h={ c_1 \over c_0} e^{-(V_1-V_0)T/2}\ .
 \end{equation}
 We shall see that the contamination by excited states needs to  be
minimised so that $h \ll 1$, while, in order to retain a reasonable 
signal to noise, we should like to achieve this with $T=2$. In order to
calibrate  the contamination by excited states we show in Table~1 the
results for the  effective potential defined as 
 \begin{equation}
 V_{\rm eff}(R,T)= -\log[W(R,T)/W(R,T-1)].
\end{equation}
 We measure the generalised correlation to $T=4$ for $R>3$ and to  $T=6$
for $R \le 3$. The signal is too noisy at larger $T$-values to  be a
useful guide.  We use a variational method with a combination of paths 
with different fuzzing levels chosen so that the effective  potential is
minimised for $T=1$. This combination is then used at all  $T$ and gives
the effective potential values shown in Table~1. As an  estimate of the
true ground state potential energy one could use  the $T=4$ effective
potential, however, this is an upper limit because of excited state
contributions. Another estimate can be made by assuming~\cite{su2pot} 
that the change in effective energy from $T=3$ to 4, namely $ \Delta V=
V_{\rm eff}(3:2)-V_{\rm eff}(4:3) $, is given entirely by the
contribution of one  excited state with an energy  difference $V_1-V_0$
determined  by the variational analysis outlined  above. Defining 
$\lambda = e^{-(V_1-V_0)}$, this  extrapolated  estimate will lie below 
$V_{\rm eff}(3:2)$ by $\Delta E /(1-\lambda)$ and this value  is shown 
in  Table~1 labelled as $T \to \infty$. The difference between these two
methods to extract  the ground state potential energy  gives an estimate
of the systematic error from extrapolation in $T$. The potentials are
the data points plotted in  Fig.~2.

In order to estimate the magnitude of the excited state contamination in
the state for $T=2$, we use the same approximation of one excited state
only for $T>2$.  Then one has  that $h^2 \approx  \Delta E /(1-
\lambda)^2$. The values of  $h$  shown in the Tables are for $T=2$ which
corresponds to our operator  having contamination $h$ at $t=T/2=1$,
namely one time step away. We see that  for the symmetric (A$_{1g}$)
representation, the contamination $h$  is quite small. 

 We also study potentials between static quarks with gluonic excitations
in  the E$_u$ representation. For this case~\cite{hyb} we need U-shaped
paths (actually  the combination $\, \sqcap - \sqcup$ with the extension
in the transverse spatial  direction). For this representation, we also 
need to extract the ground state,
and results from generalised Wilson
loops  are given in Table~2. Here we have used a single fuzzing level
but two different values of the transverse extent $d$  of the U-shaped
paths. As shown, the excited state contaminations are significantly 
higher for this representation than for the symmetric A$_{1g}$ case.

\begin{table}[ht]
\caption{ Potential energy values for the symmetric A$_{1g}$  case.}
\begin{center}
\begin{tabular}{c|c|l|l|l|l}
R& fuzz     & $V_{\rm eff}$& $V_{\rm eff}$& $V_{\rm eff}$  & $|h|$ \\
 &levels        & $T$~3:2  & $T$~4:3 & $T \to \infty$ &    \\ \hline
1&2,13 &  0.37333(5) & 0.37326(5) & 0.37325(5)& 0.010\\ % 0.238
2&2,13 &  0.56280(14) & 0.56247(17)& 0.56234(18) & 0.026\\ % 0.288
3&2,13 &  0.68233(24) & 0.68131(32) & 0.68088(36)& 0.046\\ %0.299
4&0,16,40& 0.77512(24) & 0.77394(34) & 0.77329(42) & 0.053\\ %0.355 
6&0,16,40&  0.93844(51) & 0.93685(80) & 0.93585(103)& 0.065\\% 0.387
8&0,16,40& 1.09154(77) & 1.08801(131)& 1.08558(180)& 0.100\\% 0.407 
\end{tabular}
\end{center}
\end{table}

\begin{table}[ht]
\caption{  Potential energy values for the gluonic excitation  with E$_u$
 symmetry. }
\begin{center}
\begin{tabular}{c|c|c|l|l|l|l}
R& fuzz & d    & $V_{\rm eff}$& $V_{\rm eff}$& $V_{\rm eff}$  & $|h|$ \\
 &level&          & $T$~3:2  & $T$~4:3 & $T \to \infty$ &    \\ \hline
1&13 &1,\ 2&  1.4102(12) & 1.3955(33) & 1.3813(65)& 0.24\\ % .492
2&13 &1,\ 2&  1.3484(11) & 1.3346(25)& 1.3218(49) & 0.23\\ % .481
3&13 &1,\ 2&  1.3237(11) & 1.3164(25) & 1.3099(43)& 0.16\\ % .467
4&16 &1,\ 4&  1.3157(9) & 1.3094(20) & 1.3018(41) & 0.17\\ %0.548
6&16&1,\ 4&    1.3569(9) & 1.3519(21) & 1.3455(44)& 0.16\\% 0.561
8&16&1,\ 4&   1.4341(11) & 1.4282(27)& 1.4200(62)& 0.18\\% 0.582
\end{tabular}
\end{center}
\end{table}

\section{Lattice evaluation of the colour fields}

To study the colour fields in the ground state  of the static
quark-antiquark at separation $R$, we need to evaluate the  difference
of the expectation value of a plaquette in that ground state  with its
value  in the vacuum:
 \begin{equation}
 f_R({\bf r})= {\langle V_0|\Box_{\bf r}|V_0\rangle  \over 
\langle V_0|V_0\rangle }  - \langle 0|\Box|0\rangle.
 \end{equation}
 \noindent This quantity  is  to be extracted  from lattice observables
using the geometry shown in Fig.~1. In terms of the  decomposition of
our operators into eigenstates of the transfer matrix,  the measured
matrix  element  of  a plaquette at $t$ in the generalised Wilson loop 
of size $R \times T$ is given by 
 \begin{equation}
\langle R_0| \Box_t  |R_T\rangle = 
c_0^2 e^{-V_0 T} \left( \langle V_0|\Box|V_0\rangle + 
 {c_1 \over c_0} [e^{-(V_1-V_0)t} + e^{-(V_1-V_0)(T-t)}]
 \langle V_1 | \Box | V_0\rangle +  \dots \right).
 \end{equation}
 This clearly highlights the central problem which is that contributions
from excited state contaminations will be much larger here than for the
determination of the total potential energy because  of the presence of
off diagonal terms (e.g. $ \langle V_1 | \Box | V_0\rangle $).  In order
to extract the required quantity $\langle V_0|\Box|V_0\rangle$,  we need
to  minimise the excited state contributions by taking  $T-t \ge 1$  and
$t \ge 1$ as discussed in the previous section. For the case when $T=2$ 
and $t=1$, the excited  state contribution will have a coefficient given
by $2h$ with $h$ evaluated  at $T=2$ -- as estimated in Tables~1 and 2. 
For example, to reduce the excited state contamination to 
{\mbox{\footnotesize{$\stackrel{<}{\sim}$}}} 10 \%, one needs $|h| <
0.05$.  The electric plaquettes are extended by one unit in the time
direction, however,  so one needs $T = 3$ to ensure a separation of at
least one unit to obtain  a sufficiently pure ground state.  We are also
able to investigate  the excited state contamination directly by
measuring the plaquette correlation for $T > 3$.

For odd $T$-values,  we evaluate the electric plaquette with centre at
$t=T/2$ and average  the magnetic plaquettes over $t=(T\pm 1)/2$. For
even $T$, we average  the electric plaquettes over positions with centre
at $t=(T\pm 1)/2$  whereas the magnetic plaquettes are at $t=T/2$.

For the sum rule study, we need the sum of plaquettes over all space in
principle. On the lattice, we sum the plaquettes  over all transverse 
space (a $16 \times 16 $ plane) for each $r_L$. The sum over $r_L$ is
then approximated by selecting the region within $\pm2$  lattice units
of the Wilson loop (i.e. from $r_L= - 2$ to $R+2$). We checked that this
approximation  did not introduce  any significant error since the
missing $r_L$ region contributes noise  and no signal to the
correlation. Because we sum all plaquettes including those adjacent to 
the generalised Wilson loop, it is not appropriate to use variance
reducing  techniques such as multi-hit.

 For our study at the centre point between the static quarks, $r_L=R/2$, 
we average the plaquettes with a corner (or side) at each value 
of the transverse distance $r_T$ along  a lattice axis.

\section{Sum rules for potentials}

We consider the static quark potential for separation $R$ which  is
defined as $V(R)$ in lattice units.  Then the colour fields  associated
with this pair of static quarks can be measured using  plaquettes of
appropriate orientation. Sum rules  have been derived to relate the sum
over all spatial positions  of these colour fields to $V(R)$ and its
derivative~\cite{cm}:   

\begin{equation}
{-1 \over b} \left( V+R {\partial V \over \partial R} \right) +
S_0= -\sum  ( {\cal E}_L + 2  {\cal E}_T  + 2  {\cal B}_T + {\cal B}_L) 
\label{TASU}
\end{equation} 
\begin{equation}
{1 \over 4 \beta f} \left( V+R {\partial V \over \partial R} \right) + 
E_0 = \sum  ( - {\cal E}_L + {\cal B}_L) 
\label{TELSU}
\end{equation} 
\begin{equation}
{1 \over 4 \beta f} \left( V-R {\partial V \over \partial R} \right) +
E_0 = \sum  ( - {\cal E}_T  +   {\cal B}_T ) \ .
\label{TEPSU}
\end{equation} 
 \noindent Here the sum is over all space. The quantities on the right
hand sides are  defined by taking correlations of appropriate plaquettes
with the generalised  Wilson loops as discussed in the Introduction.
Because the combination $-{\cal E} \pm {\cal B}$ corresponds to 
energy/action, we refer to these sum rules as `action', `longitudinal 
energy' and `transverse energy' respectively.

The conventional $\beta$-function is defined as $b=d\beta/d \ln a$ with 
$\beta=4/g^2$, where $g^2$ is  the bare lattice coupling. By considering
a lattice with different  lattice spacings in different directions, it
is possible to define  generalised $\beta$-functions. At the symmetry
point where $a_i=a$, one  such combination is independent and is given
by $U-S= \partial \beta_s / \partial \ln a_t- \partial \beta_t /
\partial \ln a_t =2\beta f$ in  the notations of Refs.~\cite{cm,karsch}.

The sum rules were derived~\cite{cm} for torelons, where there is no self-
energy associated with the  static quarks. For the more practical case 
of the interquark potential between static sources, a  self-energy  term
must be included in each of the sum rules - as shown above by terms
$S_0$ and $E_0$. This  self-energy will be independent of the
inter-quark separation $R$  and hence can be removed exactly by
considering differences of  the sum rules for two different $R$-values.
Moreover, it should be  independent of spatial orientation about the
source and hence is the same  for transverse and longitudinal energy as
shown above.  These self-energy terms arise from the contribution  of
each static quark line separately, thus we shall refer to both $S_0$ and
$E_0$ as `self-energy' even though $S_0$ contributes to the action sum rule.

Our evaluations of the sum rule contribution over all space are shown 
in Figs.~3-5 for the symmetric (A$_{1g}$) and transversely excited
(E$_u$) potentials. To isolate the contribution from the ground state in
each symmetry case, we need to evaluate the colour field sum using
plaquettes with  large separation from the operators which create and
annihilate the static potential. As discussed above, using our
variational operators,  $T=3$ provides a reasonable approximation to
this - allowing a separation  of one spacing between operator and
plaquette. It is important to explore  the effect of any possible 
contamination from  excited states as discussed above. The largest and
hence best determined  signal is for the action sum rule. For that case,
we are able 
for $R\le 3$ to determine the colour fields for $T=4$ and 5 and  our
evaluations  are also shown in the figures. These set the likely size of
the systematic error from excited state  contamination. For the A$_{1g}$
case, the agreement with higher $T$ values is good and confirms that we
have isolated the ground state by choosing $T=3$. For the E$_u$ case,
however, such a contamination for the action  sum rule  could be
relatively large. This is consistent with our estimates of the excited 
state contaminations $h$ given above.

We now discuss the evaluation of the left hand sides of the sum rules.
One obstacle  is the presence of the derivative $dV/dR$. Since only
discrete values of $R$ are measured, one way forward~\cite{letter} is 
to eliminate this derivative between the three sum rules which still
allows the generalised $\beta$-functions $b$ and $f$ to be determined.
Here we study the sum rules in a more  comprehensive way by explicitly
estimating the derivative. From our  lattice data for the static
potentials, we can find good interpolations (valid for  $1 < R \le 8$) 
 \begin{equation}
\label{VA1g}
 V(R)_{A1g}=0.562 + 0.0696 R - 0.255/R -0.045/R^2 
 \end{equation}
 \begin{equation}
\label{VEu}
  V(R)_{Eu} - V(R)_{A1g} = \pi/R -4.24/R^2 + 3.983/R^4
 \end{equation}
 These interpolations are illustrated in Fig.~2. Here we have used the
fact that the self-energy is the same for excited and ground states, and
the string model expectation that the excited state  has an energy
excitation of $\pi/R$ for large $R$. The coefficients of $R^{-2}$ and
$R^{-4}$ in these expressions are not intended to have explicit physical
interpretation - we just require expressions to interpolate  accurately
in $R$.

From these expressions, we can evaluate the left hand sides of the  sum
rules and they are compared with our results at $\beta=2.4$ for the
colour flux sum over all space  in  Figs.~3--5. We use the values of
the generalised  $\beta$-functions determined previously~\cite{letter},
namely  $b=-0.35$ and $f=0.61$ as an illustration. After choosing values for 
the self-energy terms as discussed below,  we find an excellent overall
agreement. The small discrepancies which  remain are comparable to the
systematic error expected from using $T=3$ to  evaluate the colour sums.
Indeed for the action sum rule, our evaluations  with $T=4$ and 5 are
also shown and these
indicate that the estimated  systematic error is large
enough to accommodate agreement. The slope of the  action sum rule
versus $R$ gives directly the $\beta$-function $b$. A somewhat  steeper
slope than that illustrated ($b=-0.35$) would fit better -- 
particularly if the $T=4$ data are selected. A preferred value for  this
comparison would be $b=-0.33$. This is compatible with our determination
of  Ref.~\cite{letter} which quoted $b=-0.35(2)$.

The perturbative expressions for these quantities 
in terms of the bare lattice coupling $\alpha=g^2/4\pi=1/\pi\beta$ 
for $SU(2)$ colour fields are~\cite{karsch}
 \begin{equation}
 b=-0.3715(1+0.49 \alpha+\dots ) 
\ \ \ ,\ \ \ \  f =1-1.13\alpha+ \dots \ .
  \end{equation}
 \noindent The bare value of  $\alpha$ is $0.13$, which gives next to
leading order perturbative values of $b=-0.395$ and $f=0.85$. The 
effective coupling is expected to be approximately  twice as big which
will decrease  the perturbative estimate for $f$ 
from 0.85 to $\approx 0.71$ and improve the 
agreement with our non-perturbative result. For $b$, however, such an 
increase in $\alpha$  will make the agreement even worse
by reducing $b$ from --0.395 to $\approx -0.42$. Thus a 
perturbative evaluation of $b$ is unreliable.

The $\beta$-function has also been studied non-perturbatively on a
lattice  by matching the measured potentials at two $\beta$ values.
Between $\beta=2.4$ and 2.5, this gave~\cite{phm} $b=-0.277(4)$
and, in Ref.\cite{PP}, using five values of $\beta$ in the range
2.35--2.55 resulted in $b=-0.304(5)$. The
value determined here is the derivative at $\beta=2.4$ rather  than
coming from a finite difference. The qualitative features of the two
approaches are in agreement,  however.
A non-perturbative study of scaling ~\cite{ekr} from a thermodynamic
approach  gave values of $b=-0.30$ and $f=0.66$ at $\beta=2.4$, although
systematic errors are not easy to quantify since an ansatz for the
functional dependence was  assumed.

The simplest assumption for the self-energy contributions is that they
are given by  evaluating the left hand sides, interpreting $V(R)$ as
$V(R)-V_0$ with $V_0=0.562$. We find that this is a reasonable
approximation for the action  sum rule since, with this assumption, we
have $S_0=0$ which is the value used in plotting  Fig.~3.  Using the
same  assumption for the two energy sum rules, however,  we find  that
an explicit  self-energy term  $E_0=0.10$ must be used to obtain  the
agreement shown  in Figs.~4 and 5.  This self-energy should be  the
same for longitudinal and transverse sum rules   and we find agreement
with a common expression. A non-zero value  arises naturally in the
derivation~\cite{cmsr} of these sum rules.

The satisfactory agreement between the sum over plaquette correlations 
and the sum rule expressions gives confidence that we are able to 
measure the colour field around a static quark and antiquark.  This 
opens up a study of the shape of the colour field distribution itself.

\section{Colour field distributions}

Since we have measured the plaquette expectation values in the gluonic
excited  state (E$_u$), we can explore the spatial distribution of the
colour  field and make comparison with models for the gluonic excitation
such as  string and flux-tube models. This has been undertaken  for the
ground state (A$_{1g}$) previously~\cite{bali,hay}. Comparison with 
models is most appropriate  at large $R$ where string-like behaviour is
expected. Here we are able to reach $R=8$ which corresponds to a
separation of order 1 fm. Unfortunately, for the E$_u$  symmetry state,
the  contamination from excited states (here we mean higher energy E$_u$
representation  states) is significant -- being of order 30\% (see Table
2).  As an exploratory study, we report our results using $T=3$ for the 
generalised Wilson loop, since this is the largest $T$-value with a
reasonable signal. The colour field distributions may then be
interpreted as applying to  a state which is predominantly the required
lowest energy E$_u$ state  but with some contamination from excited
states.

The full spatial distribution of different components of  electric and
magnetic fields is a considerable body of  data. Because we find  that
all components are approximately equal, we evaluate the  combinations of
colour fields corresponding to the average (the action)  and to those
differences given by the longitudinal and transverse energy. These three
combinations are also appropriate for sum rule study because they
correspond  to the three sum rules of Eqs.~(\ref{TASU}-\ref{TEPSU}). 
To get at the
essential behaviour, we specialise to the midpoint ($r_L=R/2$) to
minimise the  effect of self-energy components.  Results for the
transverse dependence of the colour fluxes for the action (S),
longitudinal and transverse energies ($E_{L,T}$) at $R=8$  are shown in 
Fig.~6 where the first excited state (E$_u$) case  is   compared with
the symmetric (A$_{1g}$) potential case. A qualitatively  similar
behaviour was found for $R=6$.

 We have also measured the longitudinal dependence of the sum over 
$r_T$ which is illustrated in Figs.~7 and 8. Here the effect of  the
self-energy contributions near the sources at $r_L=4$ units from the
centre can be clearly seen  for the combinations corresponding to
longitudinal  and transverse energy.

The interpretation of these results is facilitated by the sum rules.
Although the sum rules relate the sum of the colour field fluctuation 
over all space, for a string-like model the longitudinal dependence 
will be rather mild. The longitudinal profiles given by the contributions
to the overall spatial sum  from each longitudinal slice (fixed $r_L$
and summed over $r_T$) show  a
fairly flat distribution in $r_L$ for the action sum between the static
sources with the E$_u$ case showing a slightly more spread out
distribution than the A$_{1g}$ case.   A similar behaviour is shown by
the energy sums after subtracting self-energy components.  Thus some
intuition can be gained about the  colour field sums over transverse
space at the mid point $r_L = R/2$ from looking at  the left hand sides
of the sum rules divided by $R$.

 Consider an interquark  potential given by $V(R)=e/R+KR+V_0$, where the
coefficient $e \approx -0.25$ for the A$_{1g}$ potential (this is the
`Coulomb' coefficient) while in string models we expect $e =\pi$ for the
first excited string mode (which  corresponds~\cite{pm} to the E$_u$
case). Then the `transverse energy' sum rule [Eq.~(\ref{TEPSU})] will
relate the sum over  transverse energy fluctuations to $e/R$. This is
consistent with  the much larger  transverse energy seen  for the
excited gluonic case (Fig.~8) compared to the symmetric ground state
potential (Fig.~7).  The other two sum rules suggest that the action
and longitudinal energy will have  fairly similar contributions (at
$r_L=R/2$ when summed over $r_T$) for the excited gluonic and ground
state, since the left hand side is proportional to $2KR$ in each case. We
do  find approximately this behaviour for the integrals over $r_T$  as
shown in Figs.~7 and 8,  although, as shown in Fig.~6,   we see a
flatter $r_T$ distribution for the E$_u$ case than the A$_{1g}$ case.
This flatter distribution in $r_T$ for the excited gluonic case
corresponds to a `fatter' flux-tube for that case.

Before analysing in detail the distributions shown in Fig.~6, we need 
to estimate the size of any self-energy contamination. Since $R=8$,  the
self-energy contributions at 
$r_L=4$ units from the sources are
appropriate and  these  can be estimated from the measured plaquette
correlation at 4 units from either source in  the direction  away from
the other source. These are illustrated from  our data at $R=8$ in
Figs.~7 and 8 as  the points with $r_L=8$. The self-energy
contributions here  are clearly very small but, in this case, finite
size effects are potentially important since  the spatial length of the
lattice is only 16 units. Using instead  our data for the electric and
magnetic colour field contribution with $R=2,\ 4$ and 6, we find that 
this self-energy contribution to the measured plaquette correlation
summed over $r_T$ is less than $5\%$ of the signal at the  mid-point for
$R=8$. This implies that the self-energy contribution to the action 
distribution is small but that the energy distributions, which involve 
strong cancellations between electric  and magnetic plaquettes could
have larger self-energy contaminations.  This self-energy contribution
is  large enough  (and has the correct sign) to explain why the
transverse energy for the  A$_{1g}$ case is positive in Fig.~6 (which
contains the self-energy  component) whereas the sum rule analysis
described above suggests a small negative contribution (since $e< 0$). 

A way to eliminate the self-energy contribution completely is to  focus
on the difference of the A$_{1g}$ and E$_u$ distributions, since the
distributions of the self-energy will be  identical. Viewing Fig.~6 in
this light, there is seen to be a significant difference  in the
distribution in $r_T$ between the excited gluonic state and the   ground
state  and it is of interest  to compare this with models -- the topic of the
next section.

\section{Models for Flux-Tube Profiles}
\subsection{The model of Isgur and Paton}
A bosonic quantised string will not be consistent in 2 transverse
dimensions.  In order to render the string approach  applicable, Isgur
and Paton~\cite{IP}  proposed a flux-tube model in which the string
degrees of freedom are  reduced by considering a string made of  $N$
equally spaced masses. This discretisation of the string makes the string 
self-energy
finite and  this model then makes specific  predictions although they
do depend in detail on $N$. The salient  feature 
of the model is that the excitations
are transverse and so should be  observable in the transverse energy. 
This, indeed, is qualitatively in accord with our  results, since only the 
transverse energy
sum is greatly modified in  comparing the E$_u$ state to the A$_{1g}$
state. Furthermore, this is also seen directly from  the plot of the
transverse energy profile ($2E_T$) for the MC calculations in Fig.~6. 
In the next paragraphs it will be seen to what extent this qualitative feature
of the IP model compares with details of these energy profiles.

In the IP model, when the string is discretized into $N$ masses, each a
distance $a_{IP}=\frac{R}{N+1}$ apart, it is convenient to express the Hamiltonian 
in the form
\begin{equation}
\label{IPham}
H(IP,N)= \sum^N_{i=1} H_i(IP,N)=\sum^N_{i=1} (T_i+V_i) ,
\end{equation} 
where
\begin{equation}
T_i=\frac{p^2_{x_i}+p^2_{y_i}}{2m} \ {\rm and} \ 
\end{equation} 
\begin{equation}
\label{vip}
V_{i\not=1,N}=\frac{b_{IP} (N+1)}{2 R} 
\frac{1}{2}\left[(x_i-x_{i-1})^2+(x_i-x_{i+1})^2+y's\right]
\end{equation} 
\begin{equation}
V_{i=1,N}=\frac{b_{IP} (N+1)}{2 R} 
\left[x_i^2+\frac{1}{2}(x_i-x_{i\pm1})^2+y's\right].
\end{equation} 
Here $m$ is the mass  of each of the $N$ points and $b_{IP}$ is the {\em bare} 
string energy. The two are related by $m=b_{IP}a_{IP}$.
In this part of the paper, for clarity, there is  a notation change with
the earlier $r_L$, $r_T$ now becoming $z_i$, $\sqrt{x^2_i+y^2_i}$.

The energy profile associated with the point $z_i$ on the line connecting the
two quarks  has the form for the state $\psi_n(IP,x_1,...,x_N,y_1,...,y_N)$
\begin{equation}
\label{Eprofile}
E_i^n(N,x_i,y_i,z_i)=\int\prod^N_{j\not=i}dx_jdy_j 
\psi^*(IP)H_i(IP,N)\psi(IP).
\end{equation} 
The energy contained in a profile for a given $z_i$ is then
\begin{equation}
\label{nprof}
E^n_i(N,z_i)=\int dx_idy_i E^n_i(N,x_i,y_i,z_i).
\end{equation} 

In most of this work $N=3$, where for the IP ground state $(n=1)$,

$E^1_1(3)=E^1_3(3)=0.3445E^1_{{\rm T}}$ and $E^1_2(3)=0.3109E^1_{{\rm T}}$

and for the first excited state$(n=2)$ 

$E^2_1(3)=E^2_3(3)=0.3464E^2_{{\rm T}}$ and $E^2_2(3)=0.3071E^2_{{\rm T}}$, 

\noindent where the total energy in state-$n$ is given
by $E^n_{{\rm T}}=(na_1+a_2+a_3)/m$ with $a_i=2b_{IP}\sin(i\pi/8)$ being the 
IP eigenfrequencies. In each case, the kinetic ($K^n_i$) and potential ($P^n_i$)
energies are approximately equal e.g. 

$K^2_2=0.1762E^2_{{\rm T}}$ and $P^2_2=0.1308 E^2_{{\rm T}}$.

\noindent This shows that, in spite of the averaging over $2(N-1)$ 
coordinates $(x_i,y_i)$, the
resultant $E^n_i,T^n_i,P^n_i$ are still very much as expected from the basic
simple harmonic oscillator structure of the original $H(IP,N)$.
In general, 
\[\Delta E_{\rm IP}(n'-n,N)=E^{n'}_T(N)-E^n_T(N)=(n'-n)a_1(N)/m=\]
\begin{equation}
(n'-n) \frac{2(N+1)}{R}\sin(\frac{\pi}{2(N+1)})
\stackrel{N\rightarrow\infty}{\longrightarrow} (n'-n)\pi/R.
\end{equation}
For $N=1,3$ this is already a good approximation to the
$N\rightarrow\infty$ limit with \\
$\Delta E_{\rm IP}(n'-n,N)/\Delta E_{\rm IP}(n'-n,\infty)=0.90,0.97$
respectively. Another point to note is that the energy profile
$E_i^n(N,x_i,y_i,z_i)$ in Eq.~(\ref{Eprofile}) is proportional to $b_{IP}$, 
so that, for $r_T=0$, it can
be tuned to fit the corresponding lattice result. However, the energy contained
in a given profile $E^n_i(N,z_i)$ in Eq.~(\ref{nprof}) is independent of $b_{IP}$.

In  the lattice calculation it is the energy and action densities that are
measured i.e. basically in units of GeV/fm$^3$, whereas the IP model
gives the energy $E^n_i(N,x_i,y_i,z_i)$  distributed on a series on 
$N$ planes through the points $z_i$ i.e. in units of
 GeV/fm$^2$.  To be able to compare the two
 it is necessary, to smooth-out the $E^n_i(N,x_i,y_i,z_i)$
into neighbouring regions away from the $N$ planes.
This is most easily achieved by defining an $\bar{E}^n$ such that
\begin{equation}
\bar{E}^n(N,x,y,z)=\frac{E^n_i(N,x,y,z)}{a_{IP}}  \ \ {\rm for} \ \ 
\left[z_i-\frac{R}{2(N+1)}\right]\le z\le \left[z_i+\frac{R}{2(N+1)}\right]. 
\end{equation} 
 In the following, it is $\bar{E}^n$ that is compared with
the MC results.  Figs.~9 show -- separately for the ground state($n=1$) and 
first excited state ($n=2$) -- the energy profile predictions of the IP model with 
$R=8$ and $N$=1, 3 on the plane passing through the centre of the axis
connecting the two quarks. 
This illustrates the strong $N$ dependence referred to earlier with the $N=3$
profiles being much larger than those for $N=1$. But this is expected, since their
3-d spatial integrals have the form $E^n_{{\rm T}}=(na_1+a_2+a_3)/m$ for $N=3$
and $na'/m$ for $N=1$, where $a_i=2b_{IP}\sin(i\pi/8)$ and $a'=2b_{IP}\sin(\pi/4)$.
 On these same
figures the MC results are also plotted for the total energy and its transverse
component, since the latter shows the qualitative features expected from a
string-based model. But at this stage it is premature to compare the IP and MC
results for a given state, since -- as in the sumrule case -- the two approaches
involve different self-energies. Therefore, in Fig.~9c the profile differences
between the two states are shown. However, it is seen that the IP predictions
still do not correspond to the MC results. Here the value of the string energy
used in the IP model has been the empirical value of $b_sa^2=0.07$ - but 
there is no reason for the bare string energy $(b_{IP})$ relevant to the IP 
model to have this same value. Unfortunately, tuning $b_{IP}$ does not help,
since  the radial integrals of the IP profiles are constrained to be
$\approx a_1/Nm$ -- as discussed above -- and also the MC results
have the similar constraints due to the sum rules in 
Eqs.~\ref{TASU}--\ref{TEPSU}. Therefore, any attempt to improve the IP model
on the axis simply moves the position of the node  further from the axis.
The IP model is expected to be best for  large
values of $R$ and is the reason for making the above comparisons at $R$=8.
However, with smaller 
values of $R$, the results are still qualitatively the same with no
indication that the situation has improved as $R$ increases from 4 to 8.

 Since the IP model is quite easy to use and is also, at present, the
only model capable of dealing with excited states, it is of interest to
try to  locate a reason for its poor agreement with the MC data.  Here
we  study the transverse profile of different components of the total
energy density  at the midpoint both on a lattice and in the IP model.
In Figs.~10a,b we show the contributions  from the electric($ {\cal
E}$) and  magnetic(${\cal B}$) terms in Eq.~(\ref{ae2}) from the lattice
and from  the kinetic energy($K$) and the potential energy($P$) of the
IP model.   The self-energy contributions to these  electric  and
magnetic contributions are expected to be small as argued above.  There
does not seem to be any prospect of identifying the IP model
contributions  to the kinetic and potential energy separately with the
lattice electric  and magnetic contributions (or vice versa) even 
invoking the freedom to set the scale in the IP model by choosing the
bare string  tension. The feature of the lattice data that there is a
large cancellation between  the electric and magnetic contributions to
the energy is known  to have its origin~\cite{cmsr,rothe} in the trace
anomaly of QCD. This is not present  in a semi-classical model such as
the IP model.  One way to interpret these results is to note that the IP
flux-tube  model treats the flux-tube as smooth whereas the empirical
lattice  results show a rough colour flux trajectory. 

One suggestion is  to interpret the energy density of the IP model as
reproducing the  electric colour flux only~~\cite{Nair}. 
 %ZZZ can we reference the following? - instead of Nair?
 In fact this is already implied in the original IP work. 
 This can be justified  if the motion of the flux-tube in the model is
taken to be non-relativistic so that magnetic contributions will be
suppressed. The comparison of this approach is shown  in Figs.~10c,d.
By using the freedom to set the bare string  tension in the IP model, a
reasonable description of both the ground state  and first excited state
distributions can be achieved.  Here the dash-dot  curves correspond to
a compromise choice of bare string tension of $2b_s$, which allows  both
ground state and first excited state to be described simultaneously.

\subsection{The Dual Potential model of  Baker, Ball and Zachariasen}

The attraction of the IP model is its extreme simplicity. However, there are
other models in which flux-tube profiles  can be calculated.
In Ref.~\cite{BBZ} a model(BBZ), based on dual potentials, is 
developed that is expected to be best for large values of $R$. 
Also for technical reasons it is very difficult in this model to calculate
the transverse profiles for small values of $r_T$ without introducing 
large errors. Since the
overall scales, in both the BBZ model and in the lattice calculations,  are set
by the observed string energy, it is not the absolute magnitudes that
are of interest but more  the comparison between the shapes of the profiles.
In Figs.~11 it is seen that the two have similar ranges and, surprisingly,
the BBZ model does not seem to deteriorate significantly as $R$ decreases
from 8 to 1 lattice unit. This comparison is done more
quantitatively in Table~3 . There the ranges of the lattice and BBZ energy
profiles are fitted with the function $(A+Br_T)\exp(-Cr_T)$. For the
lattice case all the available data is used in the fit i.e. $0\leq
r_T\leq 3$ or 4 lattice units. However, for the BBZ model different
ranges of $r_T$ are considered. In I the range is approximately the same
as that covered by the MC data.  In II the range is extended by about 2
lattice units and it is seen that this leads to smaller values of $C$
for the smaller values of $R$.  In case III -- probably the most realistic -- the BBZ model is
restricted to  $r_T \geq 1.0$ i.e. those values of $r_T$ for which it
can be reliably evaluated, and also  $r_T$ values upto the maximum for
the lattice data.  It is seen that this last case does indeed correspond
quite closely to the trend shown by the lattice results.  

\begin{table}[ht]
\caption{Comparison between the inverse-ranges($C$) of the energy profiles for the  
lattice(MC) and the BBZ model of Ref.~\protect\cite{BBZ} -- 
by fitting with
the form $(A+Br_T)\exp(-Cr_T)$. The $C$ are in units of
$1/a\approx 8.3$fm$^{-1}$
\protect\newline MC : $C$ for the lattice data
\protect\newline BBZ I : $C$ for the BBZ model over the range 
$0.2,0.4 \leq  r_T \leq 3.0,4.0$
\protect\newline BBZ II : $C$ for the BBZ model over the range 
$0.2,0.4 \leq  r_T \leq 5.0,6.0$
\protect\newline BBZ III : $C$ for the BBZ model over the range $1.0 \leq 
r_T \leq 3.0,4.0$
}
\vspace{0.5cm}
\begin{center}
\begin{tabular}{cc|c|c|c|c|c|c} 
&R&8&6&4&3&2&1\\ \hline
MC& &1.19(7)  &1.60(13)&1.91(8) &2.19(8) &2.32(1) &3.64(5)\\
BBZ&I&1.05(1)  &1.26(1) &1.71(9) &2.24(1) &3.50(1) &5.23(1)\\
  &II &1.11(1)  &1.26(1) &1.55(1) &1.79(1) &2.45(1) &5.23(1) \\ 
  &III &1.13(2)  &1.33(1) &1.61(3)&1.63(6) &2.47(1)   &3.97(2)\\ 
\end{tabular}
\end{center}
\end{table}
Since the BBZ model (unlike that of IP) is not a semi-classical model
and contains explicitly the trace anomaly, it may be  possible to also make
realistic  predictions about the action. This would be of considerable interest,
since the value of the action is much larger than that of the energy
and so on the lattice can be extracted with greater accuracy to larger 
values of $r_T$ and so serve even better in  comparisons with the BBZ model.

\section{Conclusions}

We evaluated sum rules for the action, longitudinal energy  and
transverse energy for a range of $R$-values.  We conclude that all the 
sum rules at $\beta=2.4$ are consistent  with a set of generalised
$\beta$-functions with  values of $b=-0.35(2)$ and $f=0.61(3)$. These
non-perturbative values are  not in agreement with those  obtained by
the first two terms in the perturbative expression. An extension of our 
methods to larger $\beta$-values would be of interest to study this 
discrepancy as the coupling constant is decreased.

The distribution of the colour fields around a static quark and
antiquark  at separation $R$ has been explored. For the symmetric ground
state  potential (A$_{1g}$), the results are in agreement with previous 
studies. We also determine the distribution for the lowest  gluonic
excitation (E$_u$) for the first time. For this excited state, the
action distribution shows  a central node in its transverse dependence.
The general features of the energy distribution are an  enhanced
contribution in the transverse energy and a flatter  distribution with
$r_T$. These features are compared with string inspired models  and with
intuition from sum rules. Although no comprehensive model for the observed
behaviour  is currently available, two positive features emerge.
Firstly, the Isgur-Paton model \cite{IP} predicts an energy profile that has 
a very similar shape
to that of the electric field squared (${\cal E}$ ) on the lattice 
(Figs.~10c,d) and, secondly, the dual potential model of  
Baker, Ball and Zachariasen\cite{BBZ} predicts the correct trend for the shape
of the lattice energy profile even for small values of $R$, where their
model should not be applicable (Table 3).

The authors wish to thank M. Baker and J.S. Ball for supplying
details of their flux-tube profiles prior to publication and for
correspondence with V.P. Nair.
The authors also acknowledge that these calculations were carried
out at the Centre for Scientific Computing's C94 in Helsinki and the
DRAL(UK) CRAY Y-MP and J90.  This work is part of the EC Programme
``Human Capital and Mobility'' -- project number ERB-CHRX-CT92-0051.

\begin{figure}[ht] 
\vspace{14cm} % was 14  vof was -30:too low
\includegraphics{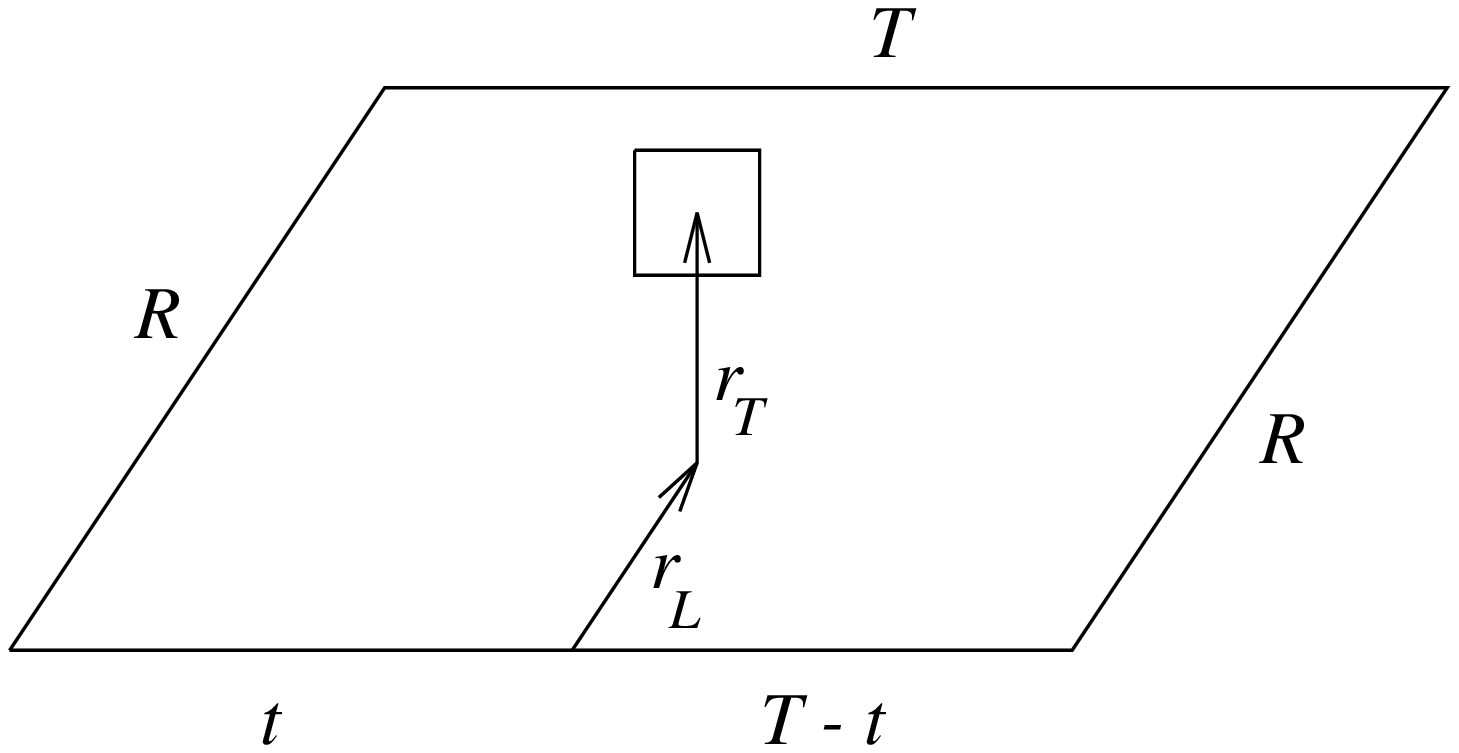}
 \caption{ The generalised Wilson loop of size $R \times T$ with 
the plaquette located at time $t$, longitudinal coordinate $r_L$ 
and transverse coordinate $r_T$.
 } 
\end{figure}
\begin{figure}[ht] 
\vspace{14cm} % was 14  vof was -30:too low
\includegraphics{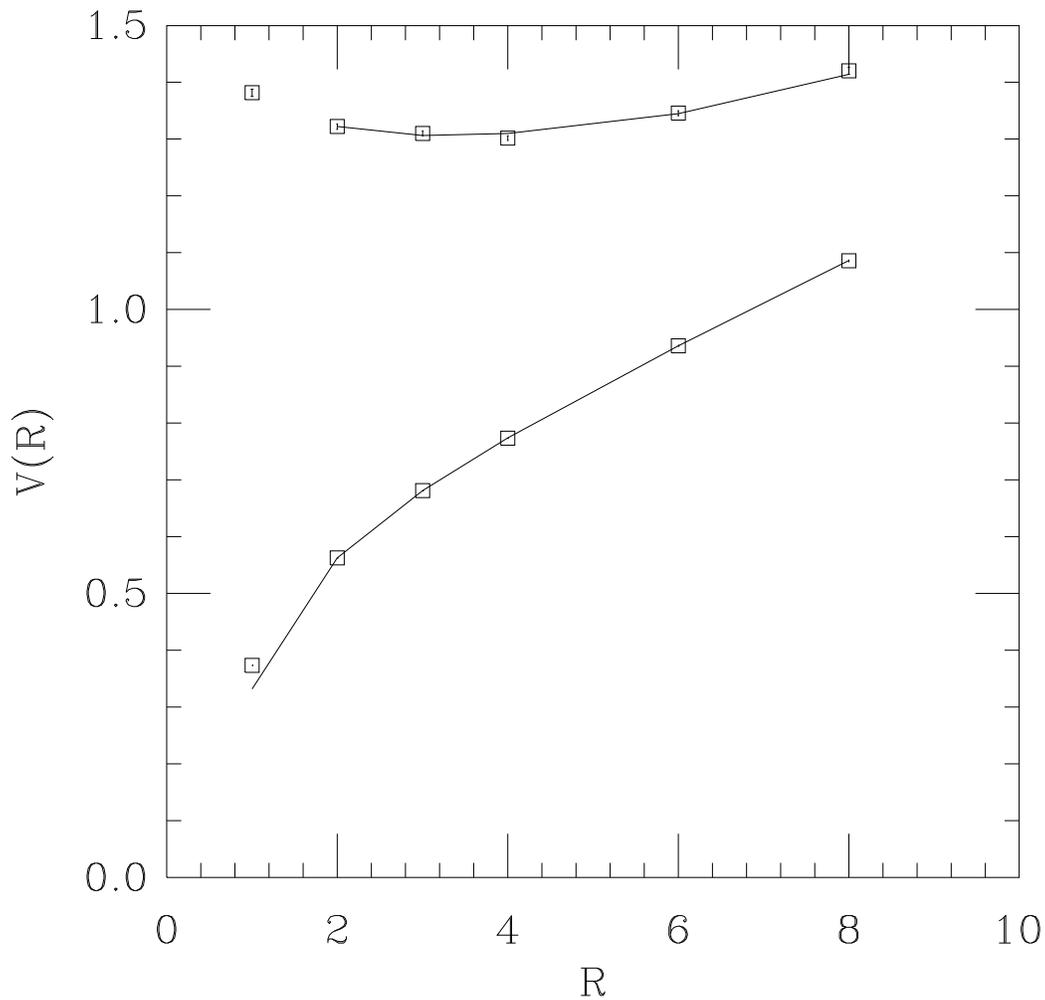}
 \caption{ The energies of the potentials in lattice units for 
separation $R$ with an interpolation
from Eqs.~(\ref{VA1g},\ref{VEu}).  The data points are for the
symmetric ground state (A$_{1g}$ representation) and first gluonic
excitation (E$_u$ representation).} 
\end{figure}

\begin{figure}[ht] 
\vspace{14cm} % was 14  vof was -30:too low
\includegraphics{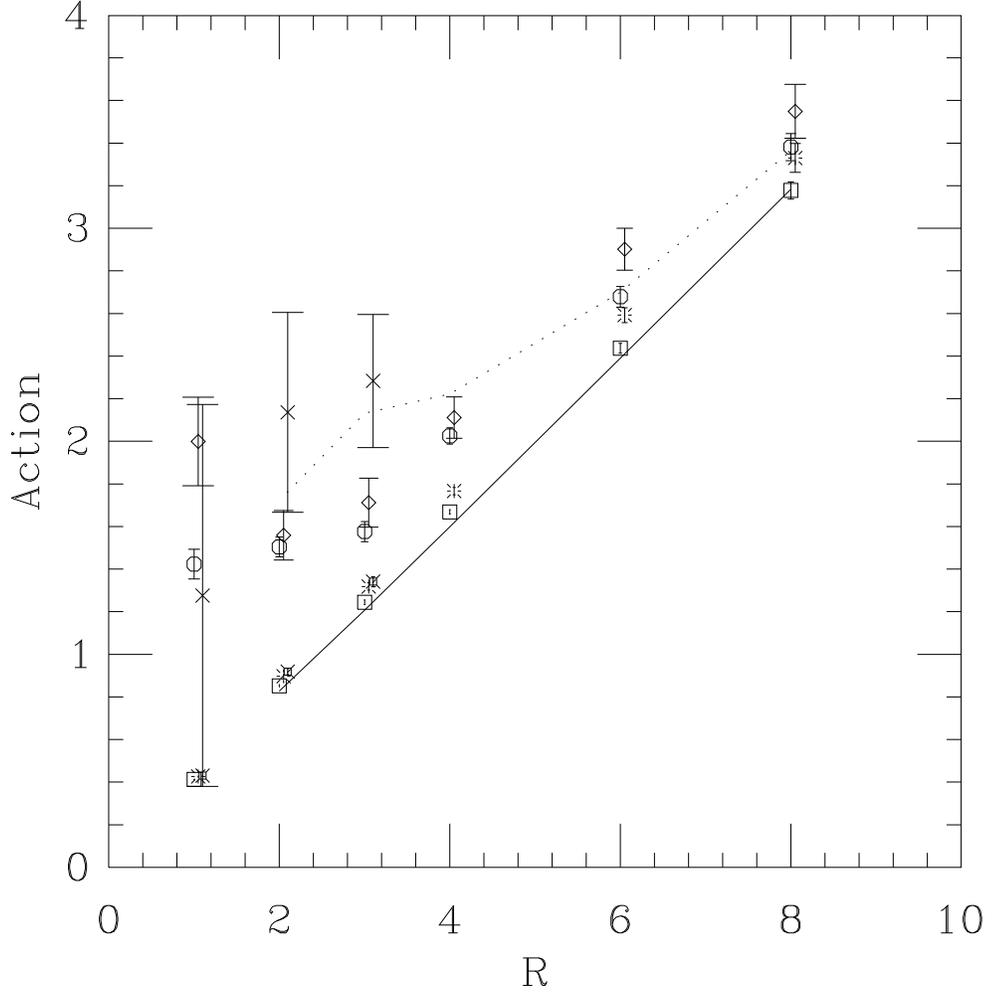}
 \caption{ The colour flux contributions in lattice units for separation
$R$ corresponding to the action ($S$) sum rule of Eq.~(12) for the static
quark potential.  The expressions for the left hand sides derived from
the  measured potentials as discussed in the text are shown by the
lines. The data points  for the symmetric ground state (A$_{1g}$
representation) are shown by squares ($T=3$),  bursts ($T=4$) and  fancy
squares ($T=5$). For the transverse gluonic excitation (E$_u$
representation), the data are shown by octagons ($T=3$), diamonds
($T=4$)  and crosses ($T=5$).
 } 
\end{figure}

\begin{figure}[ht] 
\vspace{14cm} % was 14  vof was -30:too low
\includegraphics{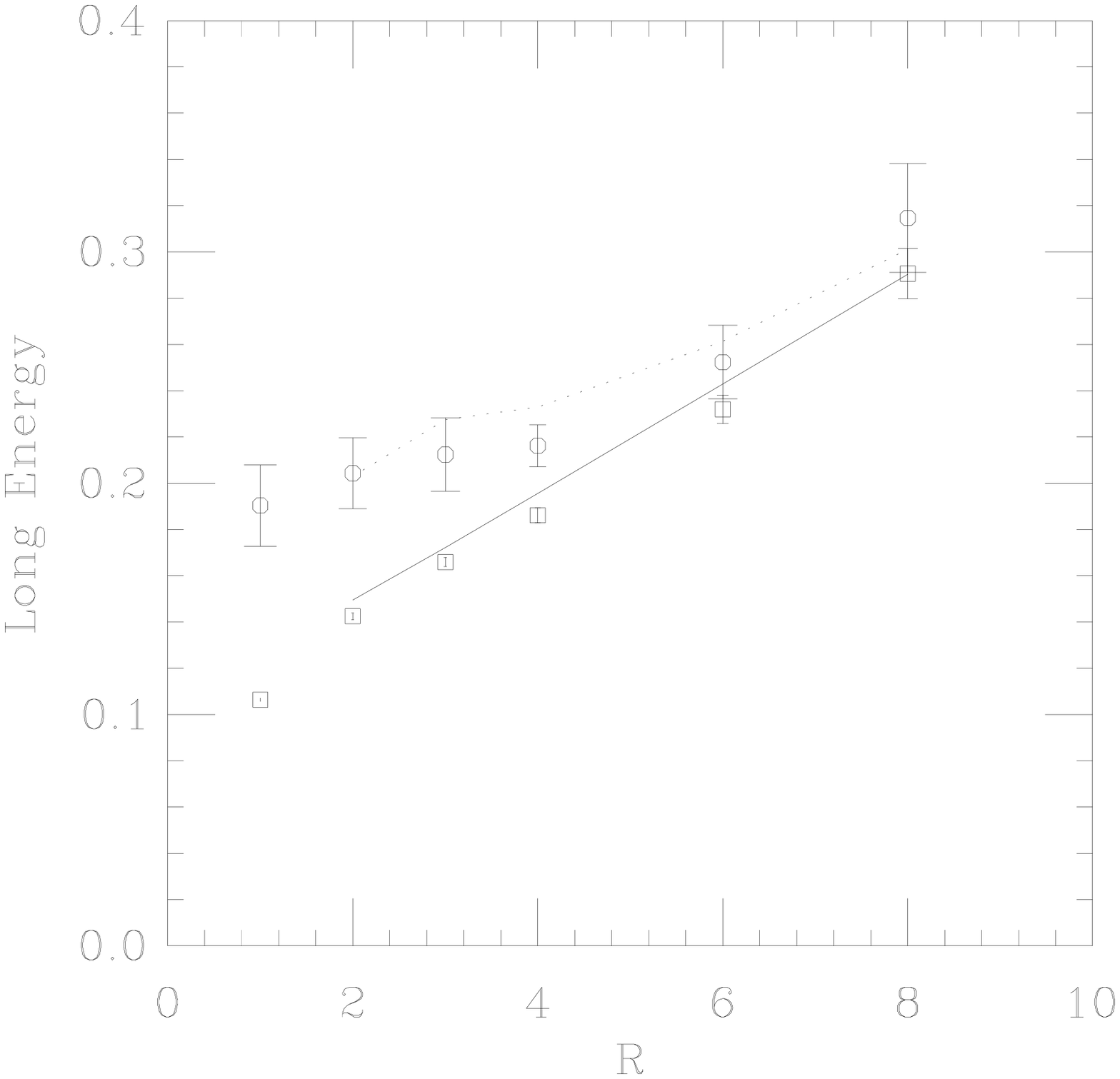}
 \caption{ The colour flux contributions in lattice units for separation
$R$ corresponding to the longitudinal ($E_L$) sum rules of Eq.~(13) for
the static quark potential.  The expressions for the left hand sides
derived from the  measured potentials are shown by the  lines. The data
points with $T=3$ are for the symmetric ground state (A$_{1g}$
representation -- squares ) and first gluonic excitation (E$_u$
representation -- octagons). 
 } 
\end{figure}

\begin{figure}[ht] 
\vspace{14cm} % was 14  vof was -30:too low
\includegraphics{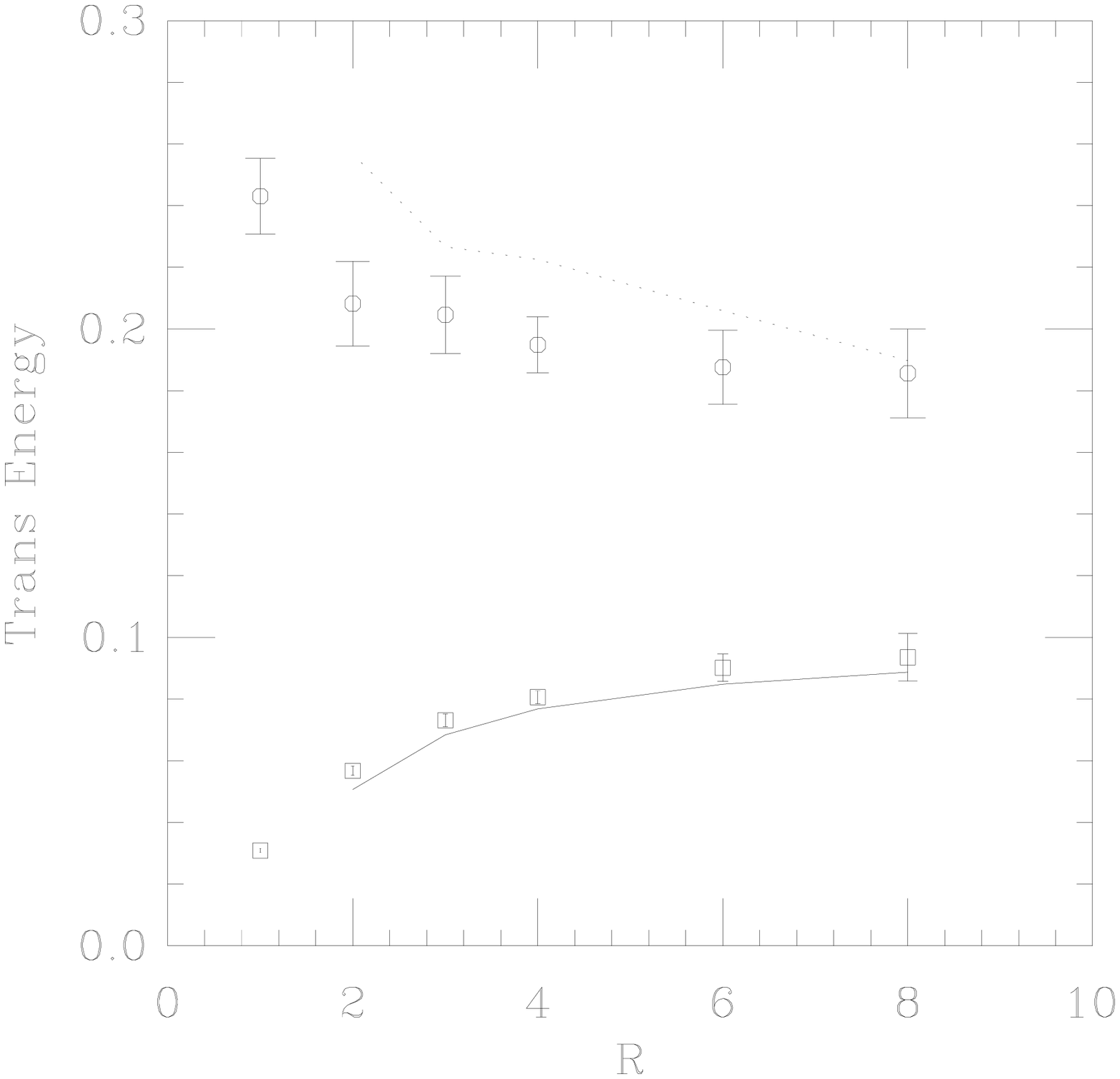}
 \caption{ The colour flux contributions in lattice units for separation
$R$ corresponding to the transverse energy ($E_T$) sum rule of Eq.~(14)
for the static quark potential.  The expressions for the left hand sides
derived from the  measured potentials are shown by the  lines. The data
points with $T=3$ are for the symmetric ground state (A$_{1g}$
representation -- squares ) and first gluonic excitation (E$_u$
representation -- octagons). 
 } 
\end{figure}

\begin{figure}[ht] 
\vspace{15cm} 
%\special{psfile=sum.ps hscale=80 vscale=75 hoffset=0 voffset=-100} 
\includegraphics{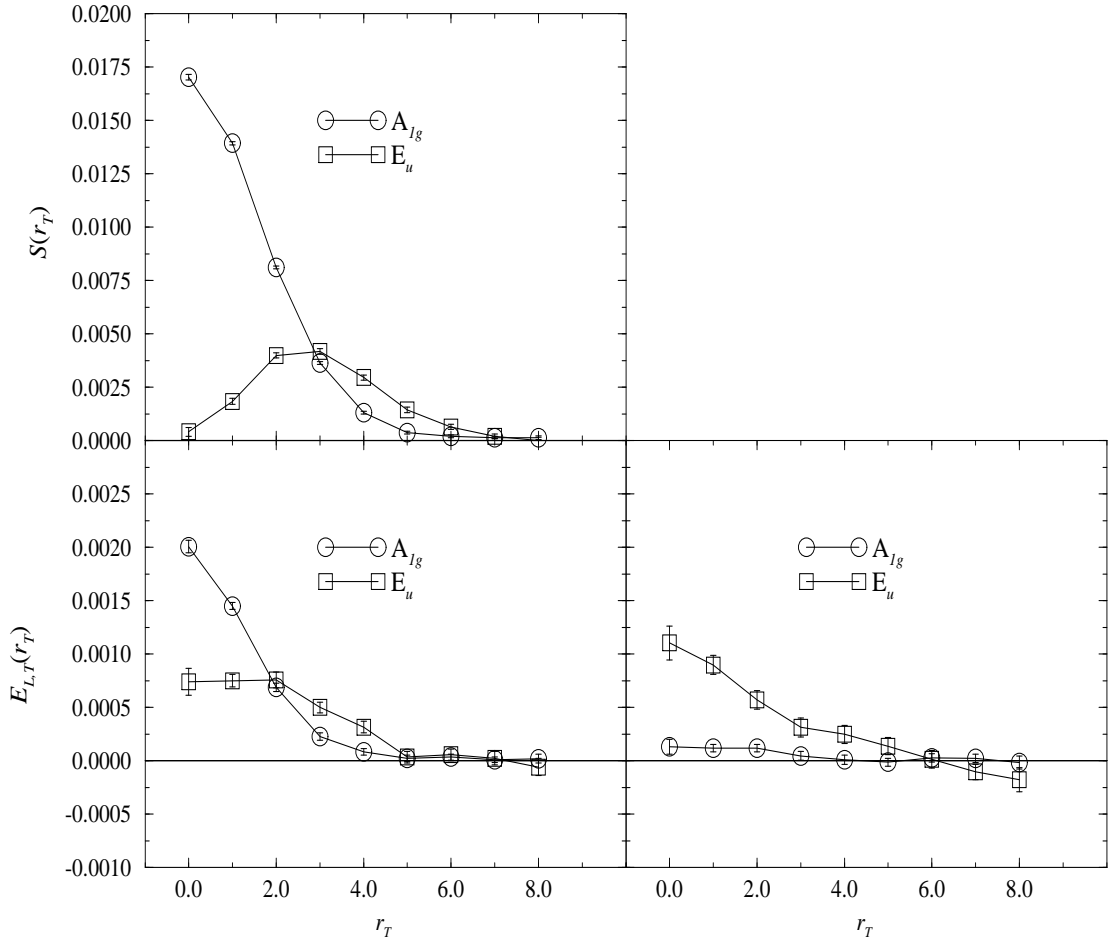} 
\vspace{2cm} 
 \caption{ The colour flux contributions corresponding to the action
($S$), longitudinal ($E_L$, left plot) and transverse energy ($2E_T$,
right plot) sum rules of Eqs.~(12--14) for the static quark potential. 
These are shown in lattice units (with $a \approx 0.6$ GeV$^{-1}$)
versus transverse distance 
$r_T$  at the mid-point ($r_L=R/2$) for
separation $R=8$.  The data are for the symmetric ground state (A$_{1g}$
representation) and first gluonic excitation (E$_u$ representation).
 } 
\end{figure}

\begin{figure}[ht] 
\vspace{14cm} 
\includegraphics{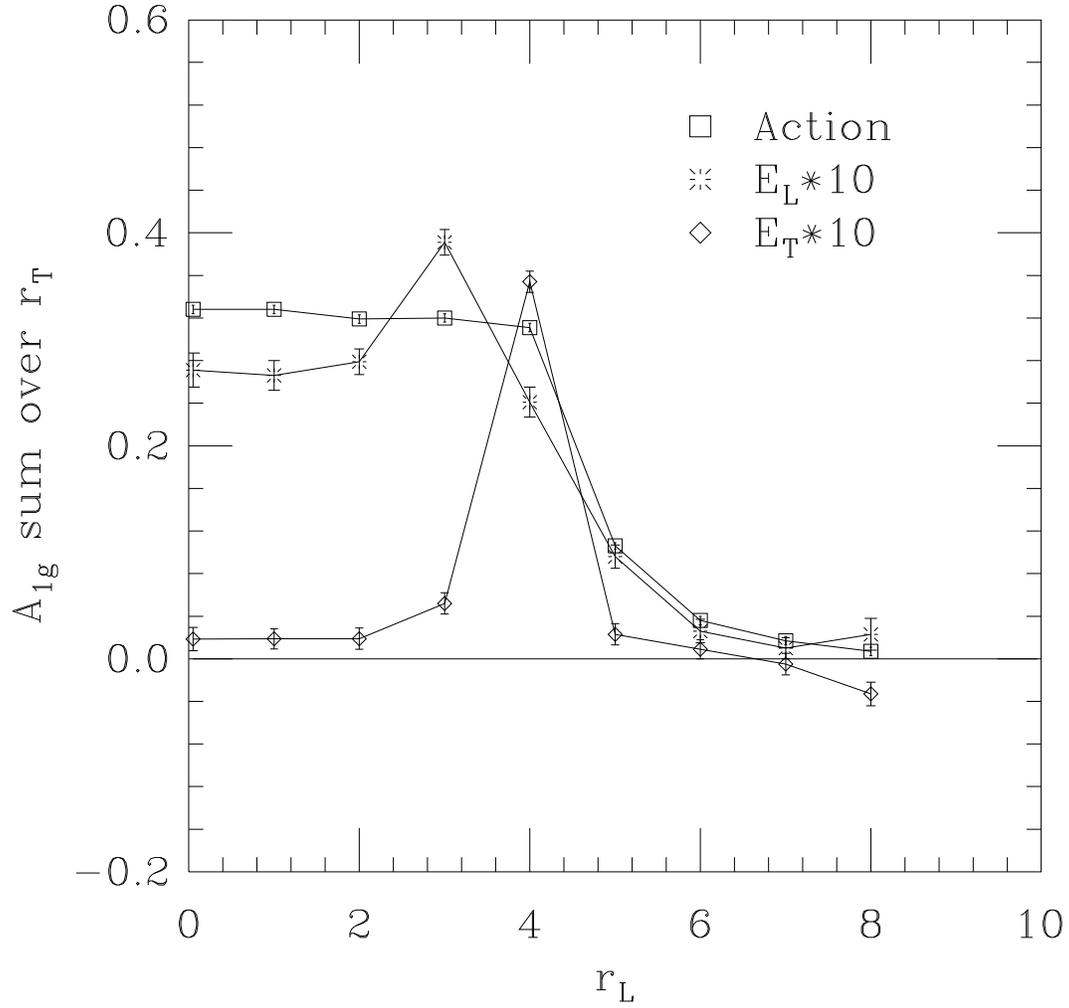}
\vspace{2cm} 
 \caption{ The dependence on longitudinal position ($r_L$) of the sum
over the transverse plane of the colour flux  contributions
corresponding to the action, longitudinal ($E_L$) and transverse
energy ($E_T$) sum rules of Eqs.~(12--14) for the static quark potential. 
Here $r_L$ is measured from the mid-point for separation $R=8$.
The data are in lattice units for the symmetric ground state (A$_{1g}$
representation).
 } 
\end{figure}

\begin{figure}[ht] 
\vspace{14cm} 
\includegraphics{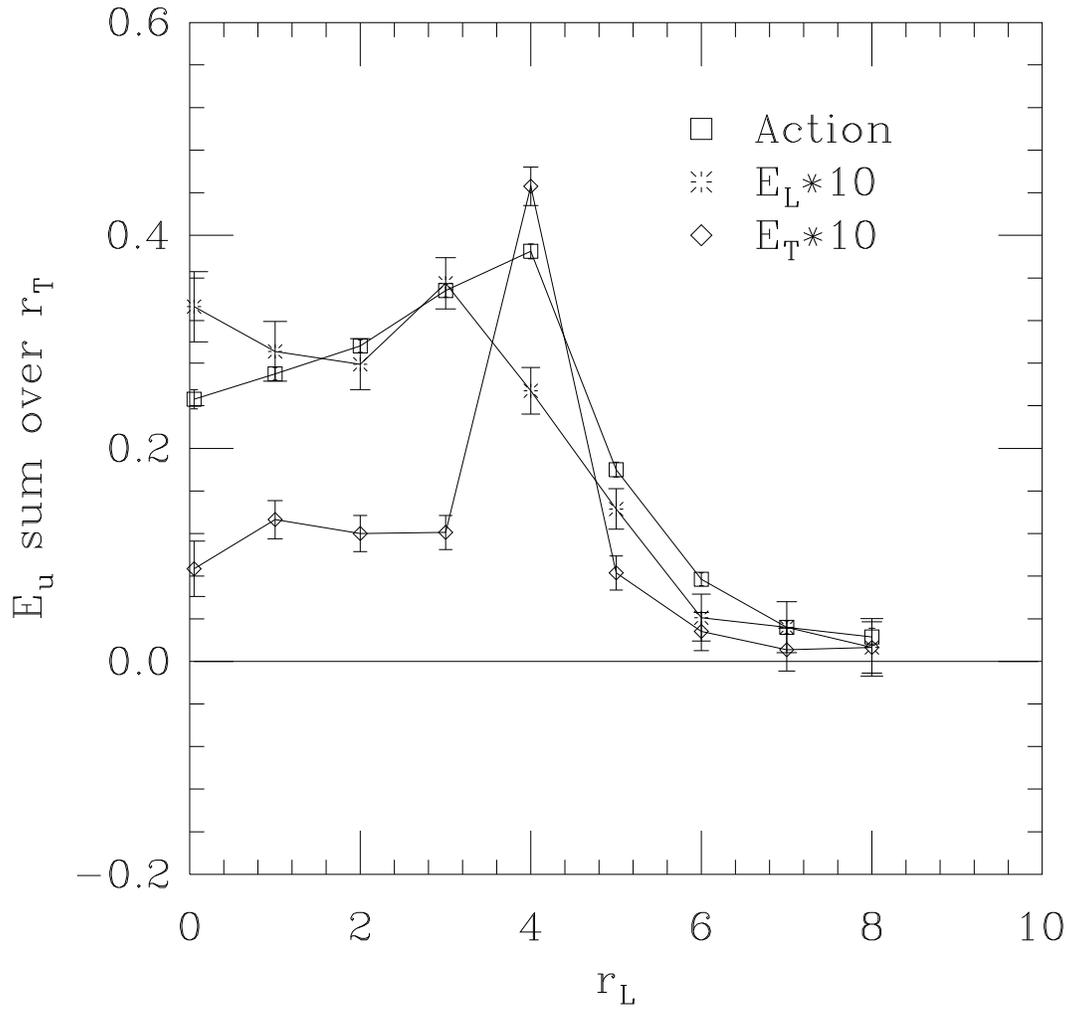}
\vspace{2cm} 
\caption{
As in Fig.~7 but for the first gluonic  excitation  (E$_u$ representation).    
 } 
\end{figure}
%\end{document}
\begin{figure}[ht] 
\vspace{14cm} 
%\special{psfile=gsmceipr8.ps voffset=-10 hscale=75 vscale=75}
\includegraphics{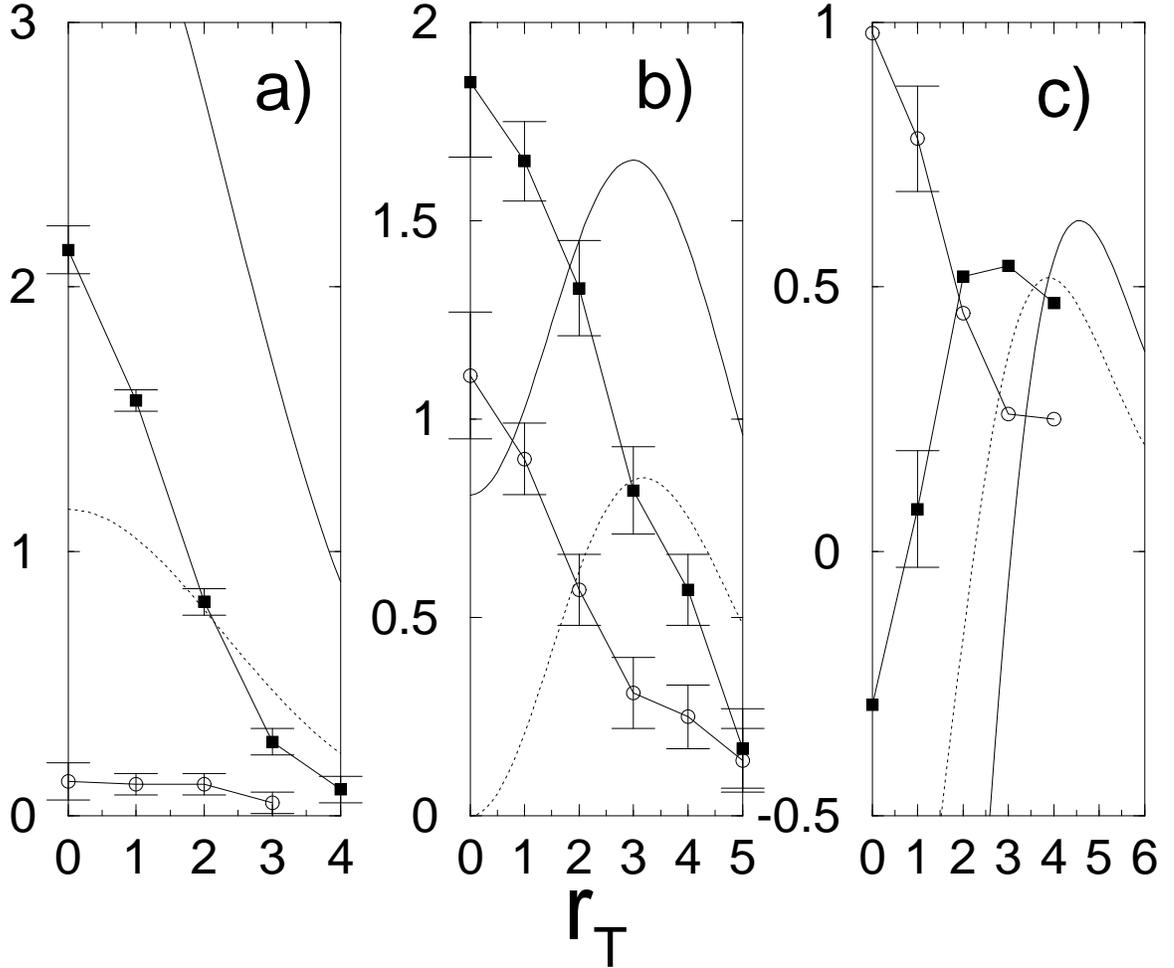}
\caption{
The energy density at the mid point versus transverse distance.  The
lattice data are at $R=8$ (using $T=3$) and are for  the Total Energy 
(solid squares) and Transverse component of the Energy (open circles).
The predictions of the IP model are for $N=1,3$ (dotted, solid). The
comparison is presented  in lattice units :
 a) For the gluonic ground state (A$_{1g}$) : 
b) For the gluonic first excited state (E$_u$) :
c) The differences (E$_u$--A$_{1g}$) between the profiles in a) and b). 
 [Compared with Figure 6 -- for convenience,  the profiles(in lattice units 
with $a \approx 0.6$ GeV$^{-1}$)
 have been multiplied by a factor of $10^3$]}
\end{figure}

\begin{figure}[ht] 
\vspace{14cm} 
%\special{psfile=tpgsmcipr8.ps voffset=-10 hscale=75 vscale=75}
\includegraphics{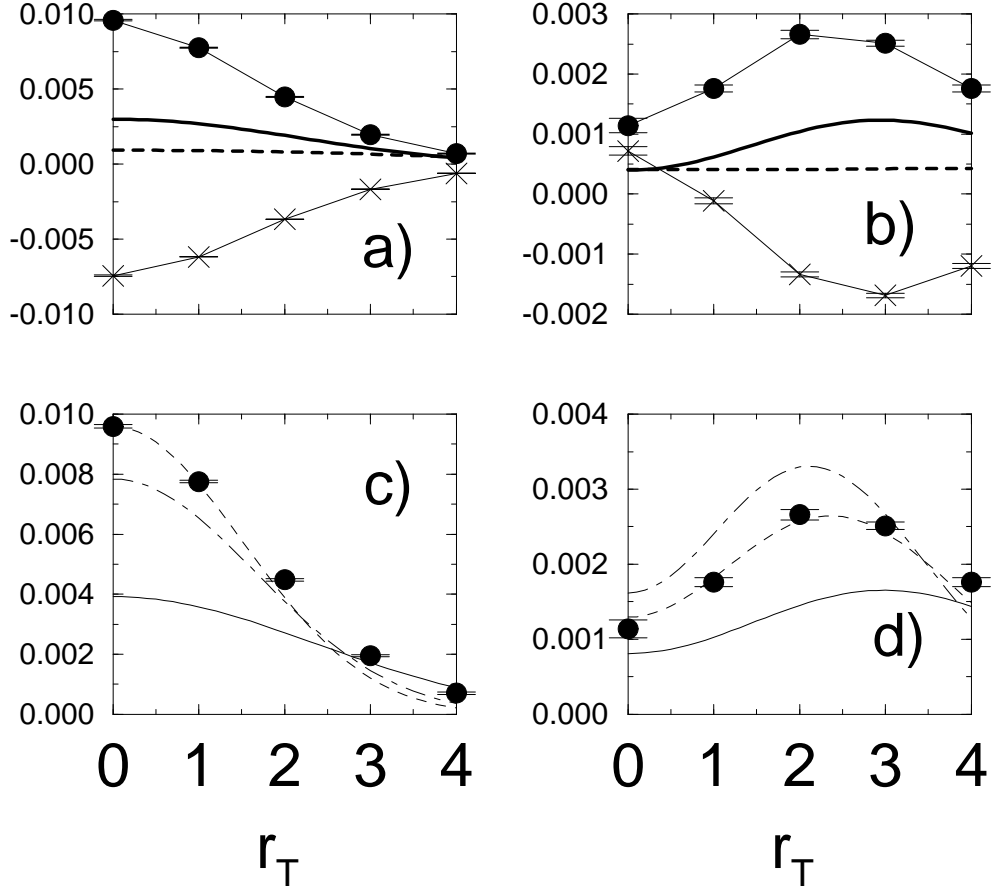}
\vspace{0cm} 
\caption{
Contributions to the energy density at the mid point versus transverse
distance. 
Comparison at $R=8$ between the $T=3$ lattice data for the 
Colour Electric Field (solid circles), Magnetic field (crosses) and the IP
predictions for the Kinetic Energy (solid), Potential Energy (dashed):
a) For the gluonic ground state (A$_{1g}$) :
b) For the gluonic first excited state (E$_u$); and 
comparison between the Colour Electric Field (solid circles) and the Total
Energy for the IP model -- using $b_{IP}=b_s=0.07$ (Solid line) , $b_{IP}=2b_s$
(Dash-dotted) and $b_{IP}$ tuned to fit lattice data (Dashed line)
c) For the gluonic ground state (A$_{1g}$)  tuning factor 2.45 for the 
Dashed line
d) For the gluonic first excited state (E$_u$) tuning factor 1.6 for the
Dashed line.
All profiles are in lattice units with $a \approx 0.6$ GeV$^{-1}$.
} 
\end{figure}
\begin{figure}[ht] 
\vspace{14cm} 
\includegraphics{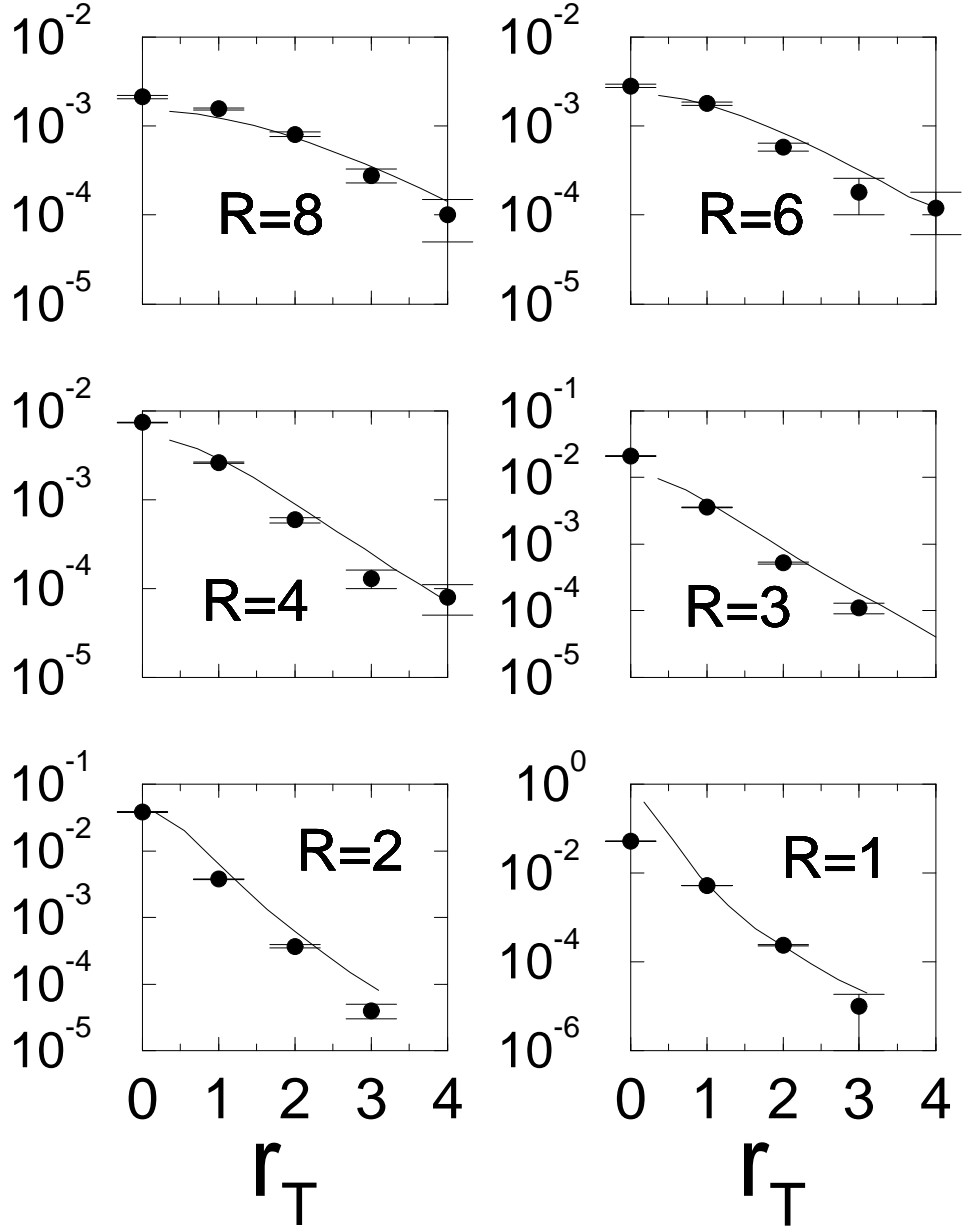}
\vspace{2cm} 
\caption{
On a semi-log scale, a  comparison of the Total Energy profile for 
the BBZ model of Ref.~\protect\cite{BBZ} (solid line) and the lattice
predictions (solid circles) for $R$ ranging from 8 down to 1 lattice unit.
All profiles are in lattice units with $a \approx 0.6$ GeV$^{-1}$.
 } 
\end{figure}

\end{document}